\documentstyle[sprocl,epsf]{article}
\def\lsim{\mathrel{\rlap{\lower4pt\hbox{\hskip1pt$\sim$}}
    \raise1pt\hbox{$<$}}}         
\def\gsim{\mathrel{\rlap{\lower4pt\hbox{\hskip1pt$\sim$}}
    \raise1pt\hbox{$>$}}}         

\def\be{\begin{equation}}
\def\ee{\end{equation}}
\def\bea{\begin{eqnarray}}
\def\eea{\end{eqnarray}}

\begin{document}
\title{NON-FACTORIZABLE CONTRIBUTIONS TO THE LARGE RAPIDITY GAP AT 
HERA$^{\star}$}

\author{  R. FIORE$^{\dagger}$ }
\address{Dipartimento di Fisica, Universit\`a della Calabria,\\ 
Istituto Nazionale di Fisica Nucleare, Gruppo collegato di Cosenza,\\ 
Arcavacata di Rende, I-87030 Cosenza, Italy}

\author{  L. L. JENKOVSZKY$^{\ddagger}$ }
\address{Bogoliubov Institute for Theoretical Physics, Academy of Sciences 
of the Ukrain, \\Kiev, Ukrain}

\author{  F.PACCANONI$^{\ast}$ }
\address{Dipartimento di Fisica, Universit\`a di Padova,\\  Istituto Nazionale 
di Fisica Nucleare, Sezione di Padova, via F. Marzolo 8,\\ I-35131 Padova, 
Italy}

\author{  E.PREDAZZI$^{\diamond}$ }
\address{Dipartimento di Fisica, Universit\'a di Torino,\\ Istituto Mazionale 
di fisica Nucleare, Sezione di Torino, via P.Giuria 1,\\ I-10125 Torino, 
Italy}

\maketitle\abstracts{ 
The large rapidity gap events from HERA are analyzed 
within a model containing a pomeron and an $f-$ reggeon contribution. The 
choice for the pomeron contribution is based on the Donnachie-Landshoff model. 
The dependence of the "effective intercept" of the pomeron on the momentum 
fraction $\beta$ and on its Bjorken variable $\xi$ is calculated.}

$ \begin{array}{ll}
^{\star}\mbox{presented at the workshop DIQUARKS 3 }\\
   \mbox{ Torino (Italy), October 28-30~~1997}
\end{array}
$
\vskip 0.5cm
$ \begin{array}{ll}
^{\dagger}\mbox{{\it email address:}} &
   \mbox{FIORE~@CS.INFN.IT}
\end{array}
$

$ \begin{array}{ll}
^{\ddagger}\mbox{{\it email address:}} &
 \mbox{JENK~@GLUK.APC.ORG}
\end{array}
$

$ \begin{array}{ll}
^{\ast}\mbox{{\it email address:}} &
   \mbox{PACCANONI~@PADOVA.INFN.IT}
\end{array}
$

$ \begin{array}{ll}
^{\diamond}\mbox{{\it email address:}} &
   \mbox{PREDAZZI~@TO.INFN.IT}
\end{array}
$
\vskip 1.5cm

	The increasing precision of the HERA-measurements and the extension of 
the kinematical domain where diffractive deep inelastic scattering (DIS)
was measured, necessitates the perfection of the relevant theoretical 
calculations beyond the simple, single, factorizable pomeron exchange, as it
was originally introduced in ~\cite{IS} and was used in subsequent papers (for 
a partial list of recent papers on the subject see ~\cite{GAP}). Although the 
role of other than a single pomeron contribution to the rapidity gap has been 
realized by the theorists long ago ~\cite{DL}, it was only recently ~\cite{D} 
that the effect was measured experimentally.

	Studies of diffractive DIS beyond the simple pole exchange 
approximation have two, albeit interrelated aspects: one is rather technical 
and has to deal with various contributions that break factorization; the other 
one is conceptual, dealing with the nature and in particular the internal 
structure of the pomeron(s).

	The pomeron itself appears to be a complicated object; according to the
perturbative QCD calculations ~\cite{L}, it corresponds to an infinite number of
singularities in the complex angular momentum plane accumulating at the
rightmost point, $J=1+\delta,\ \delta\geq 0.3.$ 

	Below we consider a simple model for the pomeron: that of Donnachie
and Landshoff ~\cite{DL}, corresponding to a single "supercritical" 
Regge pole exchange. Other options are possible, for example 
the dipole pomeron (DP) model (see ~\cite{DP}) with two terms - 
one constant and the other one rising with
energy logarithmically. This DP model is close to that of D-L as to the 
numerical fits - both producing a moderate ("soft") energy dependence, but 
the latter one may be used also as a polygon for studying the effects coming 
from the pomeron non-factorizability. Here we will limit ourselves to 
consider the first choice.

	Other possible contributions are those allowed in elastic hadron (e.g. 
$pp$) scattering: the odderon, $f, \omega,$ etc reggeons, daughter
trajectories, $\pi$ exchange etc. We confine our analysis by considering two 
major contributions, namely the pomeron (single and double poles) plus an 
effective reggeon, essentially dominated by $f$. Other contributions are either 
negligibly small or they may be absorbed by the above ones (subtle 
details - like $\pi$-exchange - have not been settled in the literature even 
in the case of the much better known case of elastic scattering).

	Explicitly, the spin-averaged differential cross-section for a 
diffractive DIS, ignoring spin and the proton mass, is (see Fig. 1 for 
kinematics and notations):
\begin{displaymath}
d\sigma=\frac{(2\pi)^{-5}}{2pl}\frac{d^3\vec{l}'}{2l'^0}
\frac{d^3\vec{p}~'}{2p'^0}\sum_n\bigg (\frac{d^3\vec{k}_n}{2k^0_n}
\delta(p+l-l'-k_n-p')|T(p+l\rightarrow p'+l'+k_n)|^2\bigg ).
\end{displaymath}

\begin{figure}
\begin{center}
\leavevmode
\epsfxsize=0.45\textwidth
\epsffile{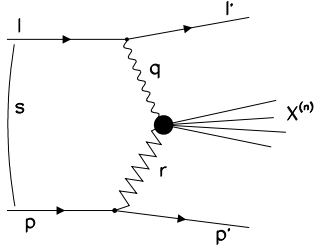}
\end{center}
\caption{Diagram for the diffractive deep-inelastic scattering.}
\label{qdsalpeter}
\end{figure}
  
	We consider the scattering amplitude corresponding to two Regge  
exchanges $R_i\ \ (R_i=P,\ f)$:
\begin{displaymath}
T\equiv T(p+l\rightarrow p'+l'+k_n)=\sum_i T^{(i)}=
\sum_iF(R_i+l\rightarrow l'+k_n)\Phi_i(\xi,t),
\end{displaymath}
where
\begin{displaymath}
\Phi_i(\xi,t)=\big (e^{i\pi/2}\xi\big )^{-\alpha_i(t)}\beta_i(t),
\end{displaymath}
and the slowly varying function $\sin(\pi\alpha/2)$ has been absorbed by the
residue.

	By denoting 
\begin{displaymath}
F(R_i+l\rightarrow l'+k_n)=\frac{e}{q^2}F(R_i+q\rightarrow k_n)=F_{R_i+l}
\end{displaymath}
we get in the case of two Regge exchanges, $P$ and $f$,
\begin{equation}
|T|^2=|F_{P+l}\Phi_P(\xi,t)|^2+|F_{f+l}\Phi_f(\xi,t)|^2+
2\Re[F_{f+l}F^*_{P+l}\Phi(\xi,t)\Phi^*_P(\xi,t)].
\ee
The first two terms in Eq.~(1) assume an immediate physical interpretation.
Really, the cross section for $R_i+l\rightarrow l'+X^{n}$ is
\begin{displaymath}
d\sigma_{R_i+l}=\frac{1}{2rl}\frac{d^3\vec{l}'}{2l'^0}
\sum_n\big (\frac{d^3\vec{k_n}}{2k^0_n}\delta^{(4)}(r+l-l'-k_n)
|F_{R_i+l}|^2\big ).
\end{displaymath}
Thus, for the $p+l\rightarrow p'+l'+X^{(n)}$ differential cross section with 
the exchange of a single reggeon $R_i$ one has
\begin{displaymath}
d\sigma^{(i)}=\frac{rl}{pl}\frac{d^3p'}{2p'^0}|\Phi(\xi,t)|^2d\sigma_{R_i+l}
\simeq\frac{\pi}{2}d\xi dt\xi|\Phi(\xi,t)|^2d\sigma_{R_i+l},
\end{displaymath}
where the relations $(rl)/(pl)\simeq(rq)/(pq)=\xi$ and
$(d^3\vec{p}~')/(2p^{'0})\simeq 1/4 d\phi'_pd\xi dt,$ valid for small
$\xi$, were used.

	The last term in the r.h.s. of Eq.~(1) is related to the imaginary 
part of the $f+l\rightarrow P+l'$ transition amplitude, and consequently, in 
$d\sigma$ a term proportional to
\begin{equation}
\frac{d^3\vec{l}'}{2l^{'0}}\sum_n\frac{d^3\vec{k}_n}{2k^0_n}
\delta^4(p+l-p'-l'-k_n)F_{f+l}F^*_{P+l}
\ee
emerges, corresponding to an $f$ meson dissociating into a pomeron. Following 
an "educated guess" by N.N.Nikolaev, W.Sch\"afer and B.G.Zakharov
~\cite{NSZ}, we ignore this contribution as being small compared to the 
elastic case.

	As recently noticed by J.Ellis and G.G.Ross ~\cite{ER}, the rapidity 
cuts commonly employed in diffractive DIS measurements require the struck 
parton in the pomeron be far off shell in a sizable region of parameter space. 
The H1 cuts correspond to very small $k^2_{min}$ that justify the treatment of 
the pomeron structure function. 
  
	Once the above-mentioned approximation has been accepted, one may write 
\begin{equation}
d\sigma_{R_i+l}=d\beta G_{q/R_i}(\beta)dq^2d\phi_{l'}\frac{d\hat\sigma}
{dq^2d\phi'_l}(k+l\rightarrow k'+l'),
\ee
in terms of the parton-lepton cross section $d\hat\sigma,$
where $k=\beta r$ and $G_{q/R_i}$ is the structure function of the
Reggeon $R_i.$  

	The above detailed presentation of the formalism was intended to fix 
the basic notions of diffractive DIS, to avoid further confusion and 
misunderstanding. 

	Now we proceed towards a quantitative analysis of the phenomenon by 
using explicit models for the reggeon fluxes and their structure functions, 
based on our experience in treating both elastic and inelastic scattering.

	Let us write our basic formula (1) in terms of the commonly used 
notation for diffractive DIS structure functions:
\begin{equation}
F_2^{D(4)}(x,t;\beta,Q^2)=A[\Phi_{q/P}(\xi,t)G_P(\beta,Q^2)+
a\Phi_{q/f}(\xi,t)G_f(\beta,Q^2)],
\ee
where, apart from the overall normalization factor $A,\ \ a$ is the only free
parameter of the model.

	The $f$-Reggeon flux is determined uniquely from standard fits to 
elastic hadron scattering (see e.g. ~\cite{LYON})):
\begin{displaymath}
\Phi_f(\xi,t)=g_f^2(t)\xi\exp{[\big(1-2\alpha_f(t)\big)L]},
\end{displaymath}
where $g_f(t)=\exp{[b_f\alpha_f(t)]},\ \ \alpha_f(t)=\alpha_f(0)+\alpha_f't$
and $L\equiv\ln {(1/\xi)}.$

We use "world average" values for the above parameters \footnote{Note that the 
$f$-reggeon intercept as extracted from various fits to the data has a trend
to increase progressively with the time. The "old-fashioned" value of
$\alpha_f(0)=0.5$ is incompatible with the data, while another extreme,
$0.8$ is claimed by some authors ~\cite{DGLM} to be compatible with the data. 
We (see e.g. ~\cite {LYON}) consider $\alpha_f(0)=0.65$ to be a reasonable 
value both in the scattering $(t<0)$ and resonance $(t>0)$ region.
Due to the strong $P-f$ mixing, the pomeron and $f$-reggeons intercepts are 
strongly correlated.} 
: $b_f=5, \ \alpha_f(0)=0.65$ and $\alpha'_f=0.9\ GeV^{-2}.$

The pomeron flux
\begin{equation}
\Phi^{D-L}_{q/P}(\xi,t)=[\exp{b(\alpha-1)}\xi^{-\alpha)}]^2\xi,
\ee
where $\alpha\equiv\alpha_P(t)=1+\delta+0.25t,\ \delta=0.08$, 
is integrated in $t$.
(For an explicit treatment of the $t-$dependence see ~\cite{GAP}c).)

\begin{figure}
\begin{center}
\leavevmode
\epsfxsize=0.55\textwidth
\epsffile{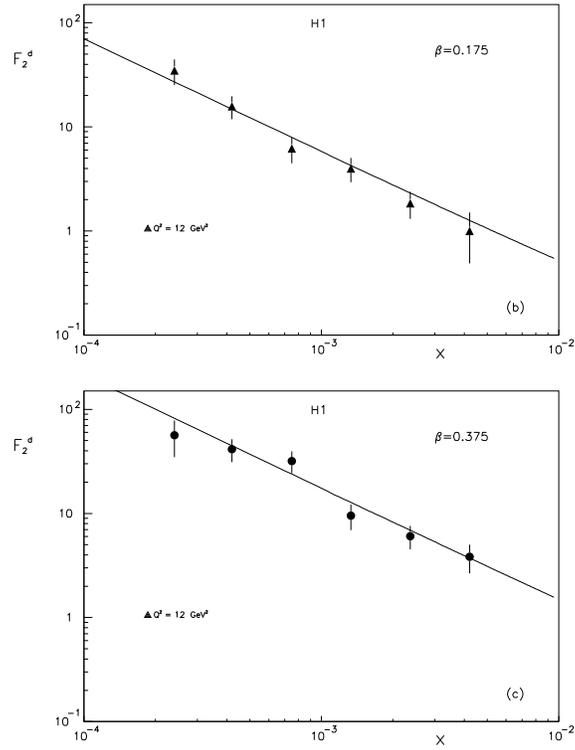}
\end{center}
\caption{A representative fit of $F_2^{D(3)}(x,\beta,Q^2)$ to the published H1 
data ~\protect\cite{H1} for fixed $Q^2=12 \,GeV^2$ and $\beta= 0.175, 0.375$.}
\label{nonfactor}
\end{figure}

We make standard choices for the structure functions. For the $f$-reggeon we
choose the following parametrization by Gl\"uck, Reya and Vogt ~\cite{GRV,D}:
\begin{equation}
\beta G_f(\beta,Q^2)=N\beta^a(1+A\sqrt\beta)(1-\beta)^D
\ee
(originally intended for pions). The values
of the parameters $N, a,\ A$ and $D,$ together with their explicit 
$Q^2$-dependence may be found in Ref.~[11]. The parameter $N$
will be rescaled by our fitting procedure. Actually, we are not too much 
concerned with any $Q^2$ dependence since it is known ~\cite{D} to be weak 
anyway.

\begin{figure}
\begin{center}
\leavevmode
\epsfxsize=0.65\textwidth
\epsffile{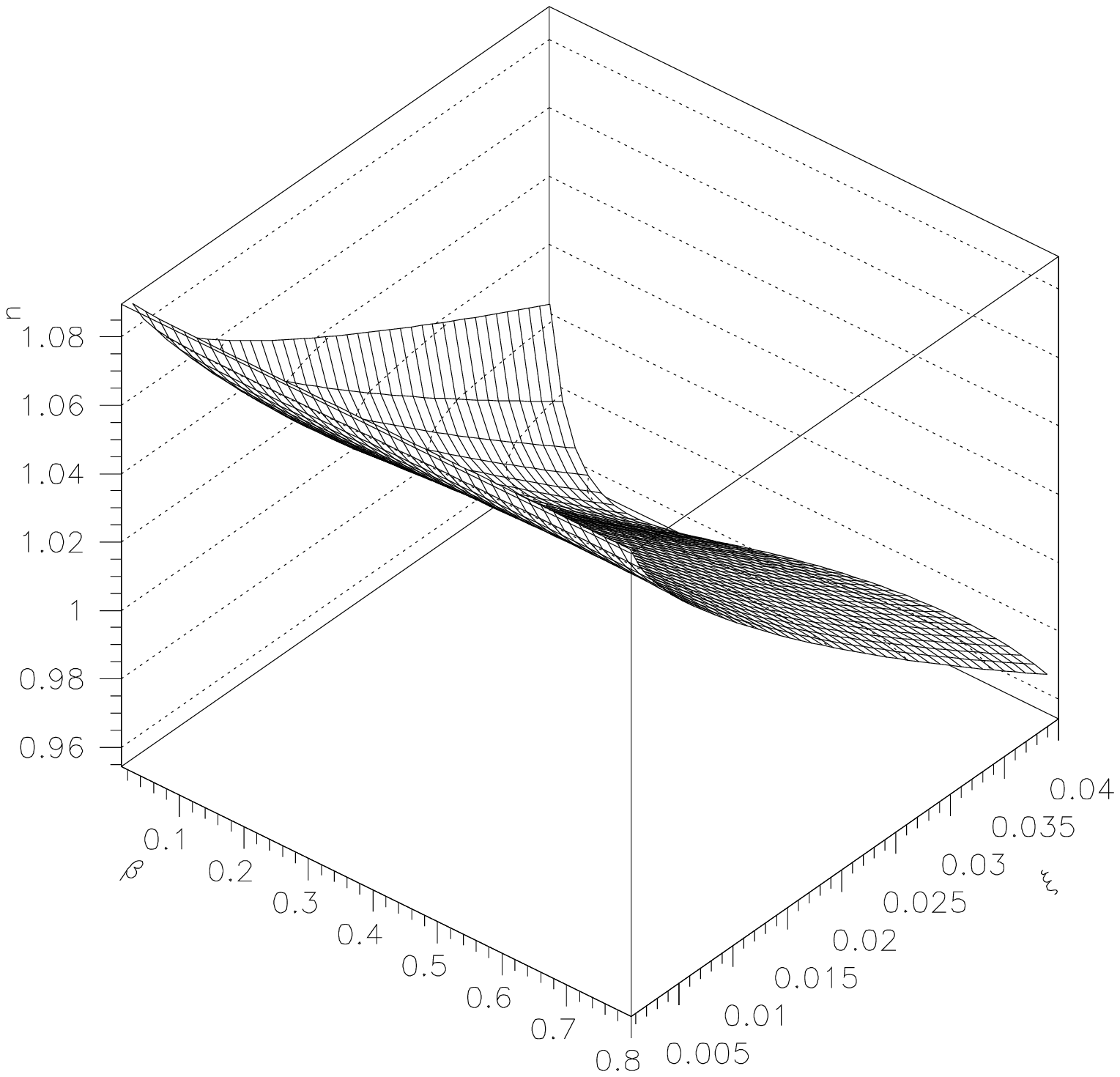}
\end{center}
\caption{$\xi$ and $\beta$ dependence of 
$n(\xi,\beta)=\partial \ln F/\partial  \ln \xi$ with the parameters fitted 
to the data of Ref.~[12]. The departure from factorization, manifest in 
the $\beta$-dependence of $n$, tends to increase with increasing $\xi$.}
\label{figy}
\end{figure}

A representative fit to the diffractive structure function $F_2^{D(3)}$
is shown in Fig. 2 for fixed $Q^2=12 \,GeV^2$ and $\beta= 0.175, 0.375$.

The non-trivial dependence of the effective power 
$n(\beta,\xi)=\partial \ln F/\partial  \ln \xi$, calculated from our model, 
is shown in Fig. 3 as a function of $\beta$ and $\xi$.
Factorization breaking, that manifests itself in the $\beta$ 
dependence of the effective power $n$,
is evident at larger $\xi$ values, where the
$f$ contribution is no more negligible with respect to the pomeron one.
Since we use in this analysis the published 1993 H1 data ~\cite{H1}, 
differences in the $\beta$-dependence of $n(\beta)$ from the result of ~\cite{D}
are expected. A maximum of $n(\beta)$, around $\beta\approx 0.6$ at 
large $\xi$, seems to be a feature present also in the 1994
preliminary H1 data ~\cite{D} where
different selection cuts, e.g. in the $t$-dependence, are applied.

To conclude, we have presented an explicit model for factorization-violating
diffractive DIS. Some details of the model may still vary but two essential
points remain invariant, namely: 1) the overwhelming contribution in the
kinematical region of diffractive DIS  as measured at HERA comes from the
pomeron and the $f$-reggeon. The
separation and identification of these objects is a complicated but at the 
same time interesting problem in doing phenomenology; 2) the pomeron emitted 
from the lower vertex in Fig.1 is the same as it
appears in elastic hadron scattering.\footnote{For the sake of definiteness, we
did not consider the case of double diffraction dissociation. Our results can be
generalized to the case when the incident proton dissociates as well.} 
Therefore, one 
can rely on pomeron models fitted to high energy $pp$ and $\bar pp$ data.
These fits unambiguously fix ~\cite{LYON} the energy dependence of the
pomeron, which is "soft". In fitting Eq.~(4) to the data, the rest of the
 parameters may be let free. 

The pomeron, which is the central object of the present analysis, itself may be
parametrized in various ways. Here, we have presented a typical model of the
pomeron, other possibilities will be considered elsewhere. Notice that a 
different type of the pomeron (see ~\cite{NSZ} and earlier references therein) 
gives rise to a strong rise of $n(\beta)$ at small $\beta$, that however may 
be compensated ~\cite{NSZ} by various subleading contributions.             


\end{document}